\documentclass[aps,prl,twocolumn,superscriptaddress,showpacs]{revtex4}
\usepackage{epsfig}
\usepackage{graphicx}
\usepackage{amsfonts}
\usepackage[figuresright]{rotating}
\usepackage{amssymb}
\usepackage{amsmath}

\def\avg#1{\langle#1\rangle}

\def\be{\begin{equation}}
\def\ee{\end{equation}}
\def\bea{\begin{eqnarray}}
\def\eea{\end{eqnarray}}

\def\nn{\nonumber}

\begin{document}
\title{Topological Insulators with SU(2) Landau Levels}
\author{Yi Li}
\affiliation{Department of Physics, University of California, San Diego,
La Jolla, California 92093, USA}
\affiliation{Princeton Center for Theoretical Science, Princeton University, Princeton, New Jersey 08544, USA}
\author{Shou-Cheng Zhang}
\affiliation{Department of Physics, Stanford University, Stanford,
California 94305, USA}
\author{Congjun Wu}
\affiliation{Department of Physics, University of California, San Diego,
La Jolla, California 92093, USA}

\begin{abstract}
We construct continuum models of 3D and 4D topological insulators by coupling
spin-$\frac{1}{2}$ fermions to an SU(2) background gauge field, which is
equivalent to a spatially dependent spin-orbit coupling.
Higher dimensional generalizations of flat Landau levels are obtained in
the Landau-like gauge.
The 2D helical Dirac modes with opposite helicities and 3D Weyl modes
with opposite chiralities are spatially separated along the third and
fourth dimensions, respectively.
Stable 2D helical Fermi surfaces and 3D chiral Fermi surfaces appear
on open boundaries, respectively.
The charge pumping in 4D Landau level systems shows quantized 4D quantum
Hall effect.
\end{abstract}
\maketitle

Time-reversal (TR) invariant topological insulators (TIs) have become a
major research focus in condensed matter physics\cite{qi2010,qi2011,Hasan2010}.
Different from the 2D quantum Hall (QH) and quantum anomalous Hall systems
which are topologically characterized by the first Chern number
\cite{klitzing1980,thouless1982,halperin1982,kohmoto1985,haldane1988}, time-reversal invariant TIs are characterized by the second Chern number in
4D \cite{zhang2001,qi2008} and the $\mathbb{Z}_2$  index in 2D and
3D\cite{kane2005,bernevig2006a,bernevig2006,qi2008,fu2007,moore2007,roy2010}.
Various 2D and 3D TIs are predicted theoretically and identified
experimentally exhibiting stable gapless
1D helical edge and 2D surface modes against TR invariant perturbations
\cite{bernevig2006,konig2007,hsieh2008,zhang2009,xia2009,chen2009}.
Topological states have also been extended to systems with particle-hole
symmetry and superconductors \cite{qi2009,kitaev2009,ryu2010}.

Most current studies of 2D and 3D TIs focus on Bloch-wave bands in lattice
systems.
The nontrivial band topology arises from spin-orbit (SO) coupling induced
band inversions\cite{bernevig2006}.
However, Landau levels (LL) are essential in the study of QH effects
because their elegant analytical properties enable construction of
fractional states.
Generalizing LLs to high dimensions gives rise to TIs
with explicit wave functions in the continuum, which would
facilitate the study of the exotic fractional TIs.
Efforts along this line were pioneered by Zhang and Hu
\cite{zhang2001}.
They constructed LLs on the compact $S^4$ sphere by coupling fermions
with the SU(2) monopole gauge potential, and various further developments
appeared \cite{karabali2002,elvang2003,bernevig2003,hasebe2010, Edge2012}.
Two-dimensional TIs based on TR invariant LLs have also been investigated \cite{bernevig2006a}.
Two of the authors generalized LLs of Schr\"odinger fermions to
high-dimensional flat space \cite{li2013} by combining the isotropic
harmonic potential and SO coupling.
LLs have also been generalized to high dimensional Dirac fermions and
parity-breaking systems \cite{li2012,li2012e}.

In all the above works, angular momentum is explicitly conserved; thus, they
can be considered as LLs in the symmetric-like gauge.
In 2D, LL wave functions in the Landau gauge are particularly intuitive:
they are 1D chiral plane wave modes spatially separated along the
transverse direction.
The QH effect is just the 1D chiral anomaly in which the chiral current
generated by the electric field becomes the transverse charge current.
In this Letter, we develop high dimensional LLs with flat spectra as
spatially separated helical Dirac or chiral Weyl fermion modes, i.e.,
the SU(2) Landau-like gauge.
They are 3D and 4D TIs defined in the continuum possessing stable gapless
boundary modes.
To our knowledge, these are the simplest TI Hamiltonians constructed
so far.
Recently, there have been considerable interests of 2D topological
band structures with approximate flat spectra.
\cite{tang2011,sun2011,neupert2011}.
Our Hamiltonians defined in the continuum possess exact flat energy
spectra in 3D and 4D,
and are independent of the band inversion mechanism.
For the 4D case, they exhibit the 4D quantum Hall effect
\cite{zhang2001,qi2008},
which is a quantized nonlinear electromagnetic
response related to the spatially separated (3+1)D chiral anomaly.
Our methods can be easily generalized to arbitrary dimensions and
also to Dirac fermions.

{\it The 3D case.---~}
We begin with the 3D TR invariant LL Hamiltonian for a spin-$\frac{1}{2}$
fermion as
\bea
H^{3D}_{LL}
&=& \frac{\vec p^2}{2m}+\frac{1}{2}m \omega_{so}^2 z^2
-\omega_{so} z (p_x \sigma_y-p_y \sigma_x),
\label{eq:3DLLHamiltonian}
\eea
which couples  the 1D harmonic potential in the $z$ direction and the 2D
Rashba SO coupling through a $z$-dependent SO
coupling strength.
$H^{3D}_{LL}$ possess translation symmetry in the $xy$ plane,
TR and parity symmetries.
Eq. (\ref{eq:3DLLHamiltonian}) can be reformulated in the form
of an SU(2) background gauge potential as
$H^{3D}_{LL}=\frac{1}{2m} (\vec{p}- \frac{e}{c}\vec{A})^2
- \frac{1}{2}m \omega_{so}^2 z^2$, where $\omega_{so}=|eG|/mc$,
and $\vec A$ takes the Landau-like gauge as $A_x=G \sigma_y z,
A_y=-G \sigma_x z$ and $A_z=0$.
In Ref. [\onlinecite{li2013}], a symmetric-like gauge with
$\vec A^\prime = G \vec \sigma \times \vec r$ is used, which
explicitly preserves the 3D rotational symmetry.
However, the SU(2) vector potentials $\vec A^\prime$ and $\vec A$ are {\it not}
gauge equivalent.
As shown below, the physical quantities of Eq. (\ref{eq:3DLLHamiltonian}),
such as density of states (DOS), are not 3D rotationally symmetric.
Nevertheless, we will see below that these two Hamiltonians give
rise to the same forms of helical Dirac surface modes, and thus
they belong to the same topological class.
A related Hamiltonian is also employed for studying electromagnetic
properties in superconductors with cylindrical geometry \cite{hirsch2013}.

Equation (\ref{eq:3DLLHamiltonian}) can be decomposed into a set of 1D harmonic
oscillators along the $z$ axis exhibiting flat spectra, a key feature of
LLs.
We define a characteristic
SO length scale $l_{so}=\sqrt{\frac{\hbar}{m \omega_{so}}}$.
Each of the  reduced 1D harmonic oscillator Hamiltonian is associated
with a 2D helical plane wave state as
$H^{z}(\vec k_{2D}) = \frac{p_z^2}{2m } + \frac{1}{2} m \omega_{so}^2
[z- l_{so}^2 k_{2D} \hat{\Sigma}_{2D} (\hat k_{2D})  ]^2$,  where $k_{2D}=(k_x^2+k_y^2)^{\frac{1}{2}}$ and $\vec k_{2D}=(k_x,k_y)$;
the helicity operator is defined as
$\hat \Sigma_{2D} (\hat k_{2D} )=\hat k_x \sigma_y -\hat k_y \sigma_x$.
The $n$-th LL eigenstates are solved as
\bea
\Psi_{n, \vec k_{2D},\Sigma}(\vec{r}) = e^{i \vec k_{2D} \cdot \vec r_{2D}}
\phi_n [z - z_0( k_{2D},\Sigma)] \otimes \chi_{\Sigma}(\hat k_{2D}),
\label{eq:3DLL_WF}
 \eea
where $\vec r_{2D}=(x,y)$; $\chi_\Sigma (\hat k_{2D})$ are eigenstates of
the helicity satisfying $\hat \Sigma \chi_\Sigma (\hat k_{2D})=\Sigma \chi_\Sigma
(\hat k_{2D})$ with helicity eigenvalues $\Sigma=\pm 1$;
$\phi_n[z- z_0]$ are the eigenstates of the $n$-th harmonic
levels with the central positions located at $z_0$,
and $z_0(k_{2D},\Sigma)= \Sigma l_{so}^2 k_{2D}$.
The energy spectra of the $n$-th LL is $E_n=(n+\frac{1}{2}) \hbar \omega_{so}$,
independent of $\vec k_{2D}$ and $\Sigma$.

\begin{figure}[htbp]
\centering\epsfig{file=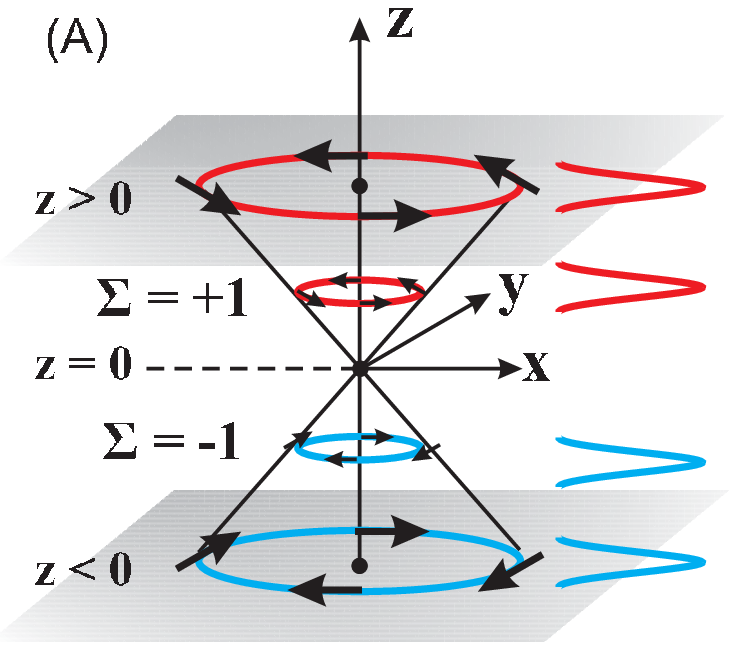,clip=1, width=0.45\linewidth,
height=0.45\linewidth, angle=0}
\centering\epsfig{file=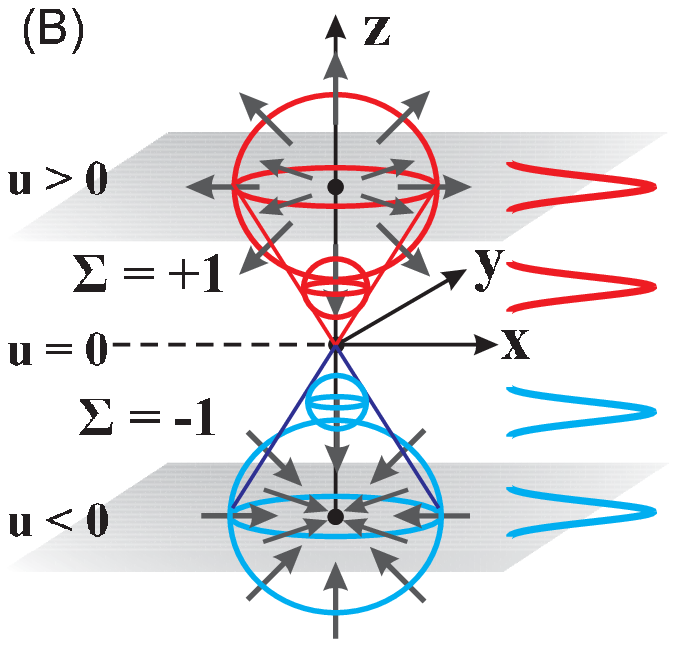,clip=2, width=0.48\linewidth,
height=0.45\linewidth, angle=0}
\caption{
3D and 4D LLs for $H^{3D}_{LL}$ and $H^{4D}_{LL}$
as spatially separated 2D helical SO plane wave modes
localized along the $z$ axis (A), and 3D Weyl modes
localized along the $u$ axis (B), respectively.
Their central locations are $z_0(\vec k_{2D}, \Sigma)=\Sigma l_{so}^2 k_{2D}$
and $u_0(\vec k_{3D},\Sigma)= \Sigma l_{so}^2 k_{3D}$, respectively.
Note that 2D Dirac modes with opposite helicities
and the 3D ones with opposite chiralities are located at
opposite sides of $z=0$ and $u=0$ planes, respectively.
}
\label{fig:3Ddemon}
\end{figure}

In the 2D LL case, spatial coordinates $x$ and $y$ are
noncommutative if projected to a given LL, say, the lowest LL (LLL).
The LLL wave functions in the Landau gauge are 1D plane waves along
the $x$ direction whose central $y$ positions linearly depend
on $k_x$ as $y_0=l_B^2 k_x$.
These 1D modes with opposite chiralities are spatially separated along
the $y$ direction.
Consequently, the $xy$ plane can be viewed as the 2D phase space of a
1D system, in which $y$ plays the role of $k_x$.
The momentum cutoff of the bulk states is determined by the system size
along the $y$ direction as $|k_x|<\frac{L_y}{2l_B^2}$.
Right and left moving edge modes appear along the upper
and lower boundaries perpendicular to the $y$ axis, respectively.
These chiral edge modes cannot exist in purely 1D systems such as
quantum wires.

Similarly, the 3D LL wave functions in Eq. (\ref{eq:3DLL_WF}) are spatially
separated 2D helical plane waves along the $z$ axis.
As shown in Fig. \ref{fig:3Ddemon} (A), for states
with opposite helicity eigenvalues, their central positions are shifted
in opposite directions.
Let us perform the LLL projection.
Among the LLL states with good quantum numbers $(\vec k_{2D},\Sigma)$,
it easy to check the following matrix elements $\avg{\Psi_{0,\vec k_{2D},\Sigma}|z|
\Psi_{0,\vec k^\prime_{2D},\Sigma^\prime}}
=\delta(\vec k_{2D}, \vec k_{2D}^\prime) \delta_{\Sigma,\Sigma^\prime}
z_0(k_{2D},\Sigma)$.
Therefore, we have
$
\avg{\Psi_{0}^a|z|\Psi^b_{0}}=\avg{\Psi_0^a|\frac{l_{so}^2}{\hbar}
(p_x\sigma_y-p_y\sigma_x)|\Psi^b_0},
$
where $\Psi_{0}^{a,b}$ are arbitrary linear superpositions of
$\Psi_{0,\vec k_{2D},\Sigma}$ in the LLL.
This proves that
\bea
P z P= \frac{l_{so}^2}{\hbar}
(p_x\sigma_y-p_y\sigma_x),
\eea
where $P$ is the LLL projection operator.
Since the LLL states span the complete basis for the plane waves
in the $xy$ plane, the projections of $x$ and $y$ in the LLL
remain themselves.
As a result, we obtain the following commutation relations
after the LLL projection as
\bea
[x, z]_{LLL}= i l_{so}^2 \sigma_y,  [y, z]_{LLL} =-i l_{so}^2 \sigma_x,
[x,y]_{LLL}=0.
\eea

Interestingly, the 3D LL states can be viewed as states in the 4D phase space
of a 2D system (with $x$ and $y$ coordinates) augmented by the helicity
structure, in which $|z|$ plays the role of the magnitude of the 2D
momentum and the sign of $z$ corresponds to $\Sigma=\pm 1$.
In fact, the momentum cutoff of the bulk states is exactly determined by
the spatial size $L_z$ along the $z$ direction as
\bea
k_{2D}<k^B_{2D} \equiv \frac{L_z}{2 l_{so}^2},
\label{2DFermi}
\eea
in which $-\frac{L_z}{2}<z<\frac{L_z}{2}$.
Applying the open boundary condition along the $z$ direction, and periodic
boundary conditions in the $x$ and
$y$ directions with spatial sizes $L_x$ and $L_y$ respectively,
we can easily count the total number of states to be
${\cal N} = \frac{L_x L_y L_z^2}{8 \pi l_{so}^4}$.
The $L_z^2$ dependence of ${\cal N}$ may seem puzzling for a 3D system,
but expressing $L_z$ in terms of Eq. ({\ref{2DFermi}), we find
${\cal N} = \frac{1}{2\pi} L_x L_y (k^B_{2D})^2$,
which is the conventional state counting of a 2D system expressed in terms of
the 4D phase space volume.
It means that an effectively 4D density of states are squeezed into
a 3D real space.

The topological property of this 3D LL system manifests through its
helical surface spectra.
Let us consider the upper boundary located at $z=z_B$ for $H^{3D}_{LL}$.
For simplicity, we only consider the LLL as an example.
If $z_B>0$, the positive helicity states $\Psi_{0,\vec k_{2D}, \Sigma=1}$ with
$k_{2D}>k^B_{2D}= z_B l_{so}^{-2}$ are confined at the boundary.
The 1D harmonic potential associated with $\vec k_{2D}$ and $\Sigma=1$
is truncated at $z=z_B$, and thus the surface spectra acquire
dispersion.
If we neglect the zero-point energy, the surface mode dispersion is
approximated by $\frac{1}{2}m\omega_{so}^2 (k-k^B_{2D})^2 l_{so}^4$ with
$k>k^B_{2D}$.
If the chemical potential $\mu$ lies above the LLL, it cuts the
spectra at surface states with a Fermi wavevector $k_f>k_{2D}^B$.
The Fermi velocity is $v_f\approx m (k_f-k_{2D}^B) \omega_{so}^2 l_{so}^4$.
The surface Hamiltonian is approximated as $H_{sf}\approx v_f (\vec p
\times \vec \sigma) \cdot \hat z -\mu$ with an electronlike Fermi
surface with $\Sigma=1$.
In another case, if the upper boundary is located at $z_B<0$, then
the negative helicity states $\Psi_{0,\vec k_{2D}, \Sigma=-1}$ with
$k_{2D}<k^B_{2D}$, and all the positive helicity states are
pushed to the boundary as surface modes.
Depending on the value of $\mu$, we can have a holelike Fermi
surface with $\Sigma=-1$,  a Dirac Fermi point, or an electronlike Fermi
surface with $\Sigma=1$.
Similarly, any other LL gives rise to a branch of gapless helical
surface modes, and each filled bulk LL contributes one helical Fermi surface.
For the lower boundary, the analysis is parallel to the above.
Each filled LL gives rise to an electronlike helical Fermi surface with
$\Sigma=-1$, or holelike with $\Sigma=1$.
According to the standard Z$_2$ classification, this system is
topologically nontrivial if the Fermi energy cuts an odd number
of Landau levels.
So far, we have assumed the harmonic frequency and SO coupling frequency
to be equal in Eq. (\ref{eq:3DLLHamiltonian}).
As explained in the Supplemental Material, although the equality of these
two frequencies is essential for the spectra flatness, the  $\mathbb{Z}_2$
topology
does not require this equality.

{\it The 4D case~}
The above procedure can be straightforwardly generalized to any higher dimension. For example, the 4D LL Hamiltonian is denoted as
\bea
H^{4D}_{LL} &=& \frac{p_u^2}{2m}+\frac{1}{2}m \omega^2 u^2
+\frac{\vec{p}^2_{3D}}{2m}-\omega u \vec{p}_{3D} \cdot \vec{\sigma},
\label{eq:4DLLHamiltonian}
\eea
where $u$ and $p_u$ refer to the coordinate and momentum of the 4th
dimension, respectively, and $\vec p_{3D}$ is the 3-momentum in the $xyz$ space.
Eq. \ref{eq:4DLLHamiltonian} can be represented as $H^{4D}_{LL}=
\frac{1}{2m} \sum_{i=1}^4 (p_i- \frac{e}{c} A_i)^2 -m \omega^2 u^2$,
where the SU(2) vector potential takes the Landau-like gauge
with $A_i=G \sigma_i u$ for $i=x,y,z$ and $A_u=0$.
Equation \ref{eq:4DLLHamiltonian} preserves the translational and rotational
symmetries in the $xyz$ space and TR symmetry.
Similar to the 3D case, the 4D LL spectra and wave functions are solved
by reducing Eq. (\ref{eq:4DLLHamiltonian}) into a set of 1D harmonic
oscillators along the $u$ axis as
$H^{u} (\vec k_{3D}) = \frac{p_u^2}{2m}+\frac{1}{2}m \omega^2
(u- l_{so}^2 k_{3D}
\hat{\Sigma}_{3D})^2$ where $k_{3D}= (k_x^2+k_y^2+k_z^2)^{\frac{1}{2}}$ and
$\hat{\Sigma}_{3D}= \hat{k}_{3D} \cdot \vec{\sigma}$.
The LL wave functions are
\bea
\Psi_{n, \vec{k}_{3D},\Sigma}(\vec{r},u) = e^{i \vec{k}_{3D} \cdot \vec{r}}
\phi_n [u - u_0(k_{3D},\Sigma)] \otimes \chi_{\Sigma}(\vec k_{3D}),
\label{eq:4DLL_WF}
\eea
where,
the central positions $u_0(k_{3D},\Sigma)=\Sigma l_{so}^2 k_{3D}$;
$\chi_{\Sigma}$ are eigenstates of 3D helicity $\hat \Sigma_{3D}$
with eigenvalues $\Sigma=\pm 1$.
Inside each LL, the spectra are flat with respect to $\vec k_{3D}$ and $\Sigma$.
This realizes the spatial separation of the 3D Weyl fermion modes as
shown in Fig. \ref{fig:3Ddemon} (B).
Similarly to the 3D case, after the LLL projection, we have the
relation $P u P= (\vec p \cdot \vec \sigma)l_{so}^2/\hbar$, and
then the noncommutative relations among coordinates are
\bea
[r_i, u]_{LLL}=il_{so}^2 \sigma_i, \ \ \, [r_i, r_j]_{LLL}=0,
\eea
for $i=x,y,z$.
The 4D LLs can be viewed
as states of the 6D phase space of a 3D system: increasing
the width along the $u$ direction corresponds to increasing the bulk
momentum cutoff $k_{3D}< k^B_{3D} \equiv L_u/(2l_{so}^2)$.
If the LLL is fully filled, the total number of states is given by
${\cal N} = \frac{L_x L_y L_z L_u^3}{24 \pi^2 l_{so}^6}$. Reexpressing
$L_u = 2k^B_{3D} l^2_{so}$
we find
${\cal N} = \frac{1}{3 \pi^2}L_x L_y L_z (k^B_{3D})^3$,
which is the conventional state counting of a 3D system expressed in terms of
the 6D phase space volume.
With an open boundary imposed along the $u$ direction,
3D helical Weyl fermion modes appear on the boundary.

Now let us consider the generalized 4D quantum Hall effects \cite{qi2008}
as the nonlinear electromagnetic response of ($4+1$)D LL
system to the external electric and magnetic (EM) fields, with
$\vec{E} \parallel \vec{B}$ in the $xyz$ space.
Without loss of generality, we choose the EM fields as
$\vec{E}=E \hat{z}$ and $\vec{B}=B \hat{z}$,
which are minimally coupled to the spin-$1/2$ fermion,
\bea
H^{4D}_{LL} (E,B)
&=& -\frac{\hbar^2}{2m}\nabla_u^2+\frac{1}{2} m \omega^2
\Big( u +il_{so}^2 \vec D \cdot \vec \sigma  \Big )^2, \ \ \,
\label{eq:4DLL_B}
\eea
where $\vec{D}=\vec{\nabla}-i\frac{e}{\hbar c} \vec{A}_{em}$.
Here $\vec A_{em}$ is the $U(1)$ magnetic vector potential in the Landau
gauge with $A_{em,x}=0, A_{em,y} = B x$ and $A_{em,z} = -cEt$.
We define $l_B=\sqrt{\frac{\hbar c}{eB}}$, where
$eB>0$ is assumed.

The $\vec B$ field further reorganizes the chiral plane wave states
inside the $n$-th 4D LL into a series of 2D magnetic LLs in the
$xy$ plane.
For the moment, let us set $E_z=0$.
The eigenvalues $E_n=(n+\frac{1}{2})\hbar \omega_{so}$ remain the same as before without splitting, while the eigen-wave-functions are changed.
We introduce a magnetic LL index $m$ in the $xy$ plane.
For the case of $m=0$, the eigen-wave-functions are spin polarized as
\bea
\Psi_{n,m=0}(k_y,k_z)&=&e^{ik_y y+ik_z z} \phi_{n}[u-u_0(k_z,m=0)]\nn \\
&\otimes& \chi_{m=0} [x-x_0(k_y)],
\label{eq:em4Dwf}
\eea
where $x_0=l_B^2 k_y$ and $\chi_0=[\phi_0(x-x_0), 0]^T$ is the zero mode channel of the operator $-i \vec D \cdot \vec \sigma$ with the eigenvalue of $\lambda_{0}=k_z$.
The central positions of the $u$ direction harmonic oscillators
are $u_0(k_z,m=0)=l_{so}^2 k_z$.
For $m\ge 1$, the eigenmodes of $-i \vec D \cdot \vec \sigma$ come
in pairs as
\bea
\chi_{m,\pm}[x-x_0(k_y)] =
\left(
\begin{array}{c}
\alpha_{m,\pm} \phi_m(x-x_0)\\
 \beta_{m} \phi_{m-1}(x-x_0)
\end{array}
\right),
\eea
where coefficients $ \alpha_{m, \pm} =l_B k_z \pm \sqrt{l_B^2 k_z^2+2m}$,
$\beta_{m}= -i \sqrt{2m}$, and the eigenvalues are
$\lambda_{m,\pm}=\pm \sqrt{k_z^2+2m l_B^{-2}}$.
The corresponding eigen-wave-functions are
$
\Psi_{n,m,\pm }(k_y,k_z)=e^{ik_y y+ik_z z}
\phi_{n}[u-u_0(k_z,m,\pm)] \chi_{m,\pm} [x-x_0(k_y)],
$
where the central positions
$u_0(k_z, m, \pm)=\pm l_{so}^2 \sqrt{k_z^2 + 2ml_B^{-2}}$.

For the solutions of Eq. (\ref{eq:em4Dwf}) with the same 4D LL index $n$, the
2D magnetic LLs with index $m=0$ are singled out.
The central positions of states in this branch are linear with
$k_z$, and thus run across the entire $u$ axis, while those of other
branches with $m\geq 1$ only lie in one half of the space as
shown in Fig. \ref{fig:4Dpump}.
After turning on $E_z$, $k_z$ is accelerated with time as
$k_z(t)=k_z(0)+eE_zt/\hbar$, and thus the central positions
$u_0(m=0,k_z)$ moves along the $u$ axis.
Only the $m=0$ branch of the magnetic LL states contribute to the
charge pumping which results in an electric current along the
$u$ direction.
Within the time interval $\Delta t$, the number of states
with each filled 4D LL passing a cross section perpendicular to
the $u$ axis is
$N= \frac{L_x L_y}{2\pi l_B^2} \frac{e E \Delta t L_z }{2\pi}$,
which results in the electric current density
$\frac{e}{L_xL_yL_z}\frac{dN}{dt}$.
If the number of fully filled 4D LLs is $n_{occ}$,
the total current density along the $u$ axis is
\bea
j_u= n_{occ} \alpha \frac{e}{4 \pi^2 \hbar} \vec{E} \cdot \vec{B},
\label{eq:4DQHE}
\eea
where $\alpha=e^2/(\hbar c)$ is the fine-structure constant.
This quantized non-linear electromagnetic response is in agreement with
results from the effective theory \cite{qi2008} as the 4D generalization
of the QH effect.
If we impose open boundary conditions perpendicular to the $u$ direction, the
above charge pump process corresponds to the chiral anomalies
of Weyl fermions with opposite chiralities on the two 3D boundaries,
respectively.
Since they are spatially separated, the chiral current corresponds to the
electric current along the $u$ direction.

\begin{figure}[tbp]
\centering\epsfig{file=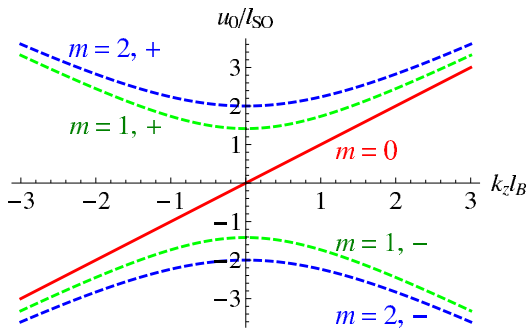,clip=1,width=0.8\linewidth, angle=0}
\caption{The central positions $u_{0}(m,k_z,\nu)$ of the 4D LLs
in the presence of the magnetic field $\vec B= B\hat z$.
Only the branch of $m=0$ (shown in red) runs across the entire $u$ axis, which gives
rise to quantized charge transport along $u$ axis in the presence
of $\vec E\parallel \vec B$ as indicated in Eq. (\ref{eq:4DQHE}).
This plot takes parameters $l_{so}=l_B$.
}
\label{fig:4Dpump}
\end{figure}

Although the DOS of our 4D LL systems is different
from the 4D TIs in the lattice system with translational symmetry
\cite{qi2008}, their electromagnetic responses obey the same Eq. (\ref{eq:4DQHE}).
It is because that only the spin-polarized $m=0$ branch of LLs,
$\Psi_{n,m=0}(k_y,k_z)$, is responsible for the charge pumping.
For this branch, $B_z$ field quantizes the motion in the
$xy$ plane so that the $x$ and $u$ coordinates play the role of
$k_y$ and $k_z$, respectively.
Consequently, the system can be viewed as the 4D phase space of
coordinates $y$ and $z$, and scales uniformly as $L_xL_yL_zL_u$
with the conventional thermodynamic limit of a 4D system.

The 3D LL Hamiltonian Eq. (\ref{eq:3DLLHamiltonian}) can be realized in strained
semiconductors by generalizing the method in Ref. \cite{bernevig2006}
from 2D to 3D.
For semiconductors with the zinc-blende structure, the $t_{2g}$ components
of the strain tensor generate SO coupling as
\bea
H_{stn}&=& -\lambda\Big\{\epsilon_{xy} (p_x\sigma_y -p_y \sigma_x)
+ \epsilon_{yz}(p_y\sigma_z- p_z \sigma_y)\nn \\
&+& \epsilon_{zx}(p_z\sigma_x- p_x \sigma_z)
\Big\},
\label{eq:strain}
\eea
where $\lambda=4\times 10^5$m/s and $9\times 10^5$m/s for GaAs and InSb,
respectively \cite{ranvaud1979}.
The $z$-dependent Rashba SO coupling is induced by the strain component
$\epsilon_{xy}$, which can be generated by applying the pressure along the
[110] direction without inducing $\epsilon_{yz}$ and $\epsilon_{zx}$
\cite{pikus1984,bernevig2006}.
Furthermore, if the pressure linearly varies along the $z$ axis,
such that $\epsilon_{xy}=gz$ where $g$ is the strain gradient,
then we arrive at the SO coupling in Eq. (\ref{eq:3DLLHamiltonian})
with the relation $\omega_{so}= g\lambda$.
A strain gradient of 1\% over the length of  $1$ $\mu$m leads
to the LL gap $\hbar \omega=2.6$ $\mu$eV in GaAs.
It corresponds to the temperature of $30$mK, which is accessible within
the current low temperature technique.
The strain gradient at a similar order has been achieved experimentally
\cite{shen1997,bernevig2006}.
The harmonic potential in Eq. (\ref{eq:3DLLHamiltonian}) is equivalent to a
standard parabolic quantum well along the $z$ axis \cite{miller1984},
which is constructed by an alternating growth of GaAs and Al$_x$Ga$_{1-x}$As
layers.
The harmonic frequency can be controlled by varying the thickness of
different layers.
}

In conclusion, we have generalized LLs to 3D and 4D in the Landau-like
gauge by coupling spatially dependent SO couplings with harmonic potentials.
This method can be generalized to arbitrary dimensions by replacing the
Pauli-matrices with the $\Gamma$ matrices in corresponding dimensions.
These high dimensional LLs exhibit spatial separation of helical or chiral
fermion modes with opposite helicities, which give rises to gapless
helical or chiral boundary modes.
The 4D LLs give rise to the quantized nonlinear electromagnetic
responses as a spatially separated (3+1)D chiral anomaly.
Many interesting open problems are left for future studies.
For example, in the Supplemental Material, we present a preliminary
effort on constructing of Laughlin wave functions in 4D LLs with spin
polarizations.
The generalizations of LLs to 3D and 4D Dirac fermions are also given
there.
A full study of the interaction effects on the high dimensional
fractional topological states based on LLs will be investigated
in future publications.

Y.L. and C.W. thank J. Hirsch, N.P. Ong, L. Sham, Z. Wang and Y.S. Wu 
for helpful discussions. 
Y.L. and C.W. are supported by the NSF DMR-1105945 and AFOSR
FA9550-11-1-0067(YIP), and S.-C. Z. is supported by the NSF under Grant
No. DMR-1305677. C.W. also acknowledges the support from the NSF of China
under Grant No. 11328403. Y.L. thanks the Inamori Fellowship and the
support at the Princeton Center for Theoretical Science.


\begin{thebibliography}{10}

\bibitem{qi2010}
X.-L. Qi and S.-C. Zhang, Phys. Today {\bf 63},  33  (2010).

\bibitem{qi2011}
X.-L. Qi and S.-C. Zhang, Rev. Mod. Phys. {\bf 83},  1057  (2011).

\bibitem{Hasan2010}
M.~Z. Hasan and C.~L. Kane, Rev. Mod. Phys. {\bf 82},  3045  (2010).

\bibitem{klitzing1980}
K. Klitzing, G. Dorda, and M. Pepper, Phys. Rev. Lett. {\bf 45},  494  (1980).

\bibitem{thouless1982}
D.~J. Thouless, M. Kohmoto, M.~P. Nightingale, and M. den Nijs, Phys. Rev.
  Lett. {\bf 49},  405  (1982).

\bibitem{halperin1982}
B.~I. Halperin, Phys. Rev. B {\bf 25},  2185  (1982).

\bibitem{kohmoto1985}
M. Kohmoto, Ann. Phys. (N. Y.) {\bf 160},  343  (1985).

\bibitem{haldane1988}
F.~D.~M. Haldane, Phys. Rev. Lett. {\bf 61},  2015  (1988).

\bibitem{zhang2001}
S. Zhang and J. Hu, Science {\bf 294},  823  (2001).

\bibitem{qi2008}
X.-L. Qi, T.~L. Hughes, and S.-C. Zhang, Phys. Rev. B {\bf 78},  195424
  (2008).

\bibitem{kane2005}
C.~L. Kane and E.~J. Mele, Phys. Rev. Lett. {\bf 95},  226801  (2005).

\bibitem{bernevig2006a}
B.~A. Bernevig and S.~C. Zhang, Phys. Rev. Lett. {\bf 96},  106802  (2006).

\bibitem{bernevig2006}
B.~A. Bernevig, T.~L. Hughes, and S.~C. Zhang, Science {\bf 314},  1757
  (2006).

\bibitem{fu2007}
L. Fu and C.~L. Kane, Phys. Rev. B {\bf 76},  045302  (2007).

\bibitem{moore2007}
J.~E. Moore and L. Balents, Phys. Rev. B {\bf 75},  121306  (2007).

\bibitem{roy2010}
R. Roy, New J. Phys. {\bf 12},  065009  (2010).

\bibitem{konig2007}
M. K\"onig {\it et~al.}, Science {\bf 318},  766  (2007).

\bibitem{hsieh2008}
D. Hsieh {\it et~al.}, Nature (London) {\bf 452},  970  (2008).

\bibitem{zhang2009}
H. Zhang {\it et~al.}, Nat. Phys. {\bf 5},  438  (2009).

\bibitem{xia2009}
Y. Xia {\it et~al.}, Nat. Phys. {\bf 5},  398  (2009).

\bibitem{chen2009}
Y.~L. Chen {\it et~al.}, Science {\bf 325},  178  (2009).

\bibitem{qi2009}
X.-L. Qi, T.~L. Hughes, S. Raghu, and S.-C. Zhang, Phys. Rev. Lett. {\bf 102},
  187001  (2009).

\bibitem{kitaev2009}
A. Kitaev, AIP Conf. Proc. {\bf 1134},  22  (2009).

\bibitem{ryu2010}
S. Ryu, A. Schnyder, A. Furusaki, and A. Ludwig, New J. Phys. {\bf 12},  065010
   (2010).

\bibitem{karabali2002}
D. Karabali and V. Nair, Nucl. Phys. B {\bf 641},  533  (2002).

\bibitem{elvang2003}
H. Elvang and J. Polchinski, C. R. Acad. Sci. {\bf 4},  405  (2003).

\bibitem{bernevig2003}
B.~A. Bernevig, J. Hu, N. Toumbas, and S.~C. Zhang, Phys. Rev. Lett. {\bf 91},
  236803  (2003).

\bibitem{hasebe2010}
K. Hasebe, Symmetry, Integrability Geom. Methods Appl. {\bf  6},    (2010).

\bibitem{Edge2012}
J.~M. Edge, J. Tworzyd\l{}o, and C.~W.~J. Beenakker, Phys. Rev. Lett. {\bf
  109},  135701  (2012).

\bibitem{li2013}
Y. Li and C. Wu, Phys. Rev. Lett. {\bf 110},  216802  (2013).

\bibitem{li2012}
Y. Li, K. Intriligator, Y. Yu, and C. Wu, Phys. Rev. B {\bf 85},  085132
  (2012).

\bibitem{li2012e}
Y. Li, X. Zhou, and C. Wu, Phys. Rev. B {\bf 85},  125122  (2012).

\bibitem{tang2011}
E. Tang, J.-W. Mei, and X.-G. Wen, Phys. Rev. Lett. {\bf 106},  236802  (2011).

\bibitem{sun2011}
K. Sun, Z. Gu, H. Katsura, and S. Das~Sarma, Phys. Rev. Lett. {\bf 106},
  236803  (2011).

\bibitem{neupert2011}
T. Neupert, L. Santos, C. Chamon, and C. Mudry, Phys. Rev. Lett. {\bf 106},
  236804  (2011).

\bibitem{hirsch2013}
J.~E. Hirsch (private communication); Journal of Supercond. {\bf 26},  2239  (2013).

\bibitem{ranvaud1979}
R. Ranvaud, H.-R. Trebin, U. R{\"o}ssler, and F.~H. Pollak, Phys. Rev. B
  {\bf 20},  701  (1979).

\bibitem{pikus1984}
G. Pikus and A. Titkov, Spin Relaxation under Optical Orientation in Semiconductors, Modern Problems in Condensed Matter Sciences Vol. {\bf 8},  Chapter 3,  (North-Holland, Amsterdam, 1984), p. 73.

\bibitem{shen1997}
Q. Shen and S. Kycia, Phys. Rev. B {\bf 55},  15791  (1997).

\bibitem{miller1984}
R.~C. Miller, A.~C. Gossard, D.~A. Kleinman, and O. Munteanu, Phys. Rev. B {\bf
  29},  3740  (1984).

\end{thebibliography}

\begin{thebibliography}{10}


\bibitem{rezayi1994}
E.~H. Rezayi and F.~D.~M. Haldane, Phys. Rev. B {\bf 50},  17199  (1994).

\bibitem{jackiw1976}
R. Jackiw and C. Rebbi, Phys. Rev. D {\bf 13},  3398  (1976).

\bibitem{niemi1986}
A.~J. Niemi and G.~W. Semenoff, Phys. Rep. {\bf 135},  99  (1986).

\bibitem{jiang2011}
Y. J. Jiang, F. Liu, F. Zhai, T. Low, and J. P. Hu, Phys. Rev. B
{\bf 84}, 205324 (2011).

\end{thebibliography}

\section{ Supplemental Material}

In this supplemental material, we cover the classic equations of motion
of the 3D Landau levels, further discussion of the topology class,
the construction of Laughlin-type wavefunction, and the 3D Landau
levels for relativistic fermions.

\section{Classical equations of motion}
The classical equations of motion for $H^{3D}_{LL}$ (Eq. (1) in the main text)
are derived as
\bea
\dot{\vec{r}}&=& \frac{\vec{p}- \frac{e}{c}\vec{A}}{m}, \nn \\
\dot{\vec{p}}_{2D}&=&0, \hspace{22mm}
\dot{p}_z=  2 \omega (\vec{p} \times \vec{S})_z- m \omega^2 z, \notag \\
\dot{\vec{ S}}_{2D}&=&
- \frac{2\omega}{\hbar} z S_z \vec{p}_{2D}, \hspace{5mm}
\dot{S}_z = \frac{2\omega}{\hbar} z \vec{S}_{2D} \cdot \vec{p}_{2D},
\eea
where $\vec S_{2D}$ and $\vec S_z$ are spin operators in the Heisenberg
representation defined as
$\vec S_{2D}=\frac{1}{2}e^{iH_{LL}^{3D}t}(\sigma_x, \sigma_y)
e^{-iH_{LL}^{3D}t}$ and $S_z=\frac{1}{2} e^{iH_{LL}^{3D}t} \sigma_z e^{-iH_{LL}^{3D}t}$.

If we choose the initial condition of spin $\vec S$ such that $S_{z,0}=0$ and
$\vec S_{2D,0} \perp \vec p_{2D,0}$, then $\vec S$ is conserved
and lies in the $xy$ plane.
The motion is reduced to a coplanar cyclotron one in the vertical plane
perpendicular to $\vec S$.
In other words, the orbital angular momentum and spin are locked, a feature
from SO coupling.

\section{Discussions on the mismatch of the trap and spin-orbit
frequencies}

In principle there should be two different frequencies in the 3D LL
Hamiltonian Eq. 1 in the main text, the harmonic frequency denoted as
$\omega_T$ below and the spin-orbit frequency $\omega_{so}$.
Below, we will explain that our $\mathbb{Z}_2$ analysis does not depend on
the exact equality between $\omega_T$ and $\omega_{so}$, and thus is generic.
This equality is only important to maintain the spectra flatness which
the usual 3D TIs do not possess.

Let us move back to a familiar example of the usual 2D LLs of quantum
Hall systems in the Landau gauge.
In a more general form, the Hamiltonian can also be represented in
terms of two independent frequencies as
\bea
H_{2D} &=& \frac{\vec p^2}{2m}+\frac{1}{2}m \omega^2_T y^2 - \omega_0 y p_x
\nn \\
&=& \frac{p_y^2}{2m} +\frac{1}{2} m \omega_T^2 (y- \alpha l_0^2  p_x)^2
+\frac{1}{2m}(1-\alpha^2) p_x^2, \ \ \
\label{eq:H_2D}
\eea
where $\alpha=\omega_0/\omega_T$ and $l_0^2=\hbar/(m\omega_0)$.
The case of $\alpha=1$ corresponds to the usual 2D LL Hamiltonian
\bea
H_{B}=(\vec P -\frac{e}{c} \vec A)^2/2m
\label{eq:2D_LL}
\eea
with $A_x=0$, $A_y=Bx$ and $\omega_0=eB/mc$.
Eq. (\ref{eq:H_2D}) can also be viewed as Eq. (\ref{eq:2D_LL})
level with an extra harmonic potential along the $y$ axis
with
\bea
\Delta H=\frac{1}{2}m (\omega_T^2-\omega_0^2) y^2.
\eea
In other words, $H_{2D}=H_B+\Delta H$.
$\omega_T$ needs to be larger than $\omega_0$, i.e. $\alpha\ge 1$,
to ensure the spectra bounded from below.
If $\alpha=1$, the spectra are exactly flat.
If $\alpha<1$, the eigenstates are still plane-wave modes along
with the good quantum number $k_x$ whose central positions
are shifted according to $y_0(k_x)=\alpha l_0^2 p_x$.
The feature of the spatially separated chiral modes does not change.
As a result, the topology remains the same as before, but their spectra
become dispersive.
If $\alpha$ is close to 1, the dispersion is very slow.
In this case, we can view this potential difference $\Delta H$ as a soft
external potential imposing on the bulk Hamiltonian.
The edge spectra of this quantum Hall system remains chiral, only the Fermi
velocities are modified.
This situation is met in mesoscopic quantum Hall systems.
The topology of such a system is still characterized by the number
of chiral edge modes.

Coming back to our case of 3D LL Hamiltonian Eq. (1) in the main text, the
situation is very similar.
As long as the time-reversal symmetry is preserved, and the energy scale
of perturbations is much smaller than the Landau level gap, the
$\mathbb{Z}_2$ topology is maintained.
The concept of helicity due to spin-orbit coupling replaces chirality in
the usual 2D Landau levels.
When $\omega_T$ and $\omega_{so}$ do not match, the Hamiltonian is expressed as
\bea
H^{3D}_{LL}
&=& \frac{\vec p^2}{2m}+\frac{1}{2}m \omega_T^2 z^2
-\omega_{so} z (p_x \sigma_y-p_y \sigma_x),\nn \\
&=& \frac{p_z^2}{2m} +\frac{1}{2} m\omega_T^2 [z-\alpha l_{so}^2
(p_x\sigma_y-p_y\sigma_x)]^2\nn \\
&+& \frac{1}{2m} (1-\alpha^2) (p_x^2+p_y^2),
\eea
which gives rise to the following dispersive spectra as
\bea
E_{n,\vec k, \pm}= (n+\frac{1}{2})\hbar \omega_T^2
+ (1-\alpha^2)\frac{\hbar^2}{2m}  (k_x^2+k_y^2).
\eea
However, the eigenstates remains spatially separated helical 2D plane
wave modes with good quantum numbers of $(k_x,k_y,\Sigma)$,
whose central positions are shifted according to $z_0(\vec k,\Sigma)
=\alpha \Sigma l_{so}^2 k$.
As a result, the topology remains the same as before.
In other words, the mismatch between $\omega_T$ and $\omega_0$
is equivalent to add the bulk Hamiltonian Eq. (1)
with an external potential
\bea
\Delta H(r)=\frac{1}{2}m (\omega_T^2-\omega_{so}^2) z^2
\eea
In the case of $\alpha$ close to 1, this soft external potential imposes
a finite sample size along the $z$ axis with the width
$W^2<\hbar /(m\Delta \omega)\approx l_{so}^2 \omega_T/\Delta \omega$
where $\Delta\omega=\omega_T-\omega_0$.
Inside this region, $\Delta H$ is smaller than the LL gap, and the LL
states are bulk states.
The corresponding 2D wavevectors for these bulk states satisfy
$k^2_{2D}<l_{so}^{-2} \frac{\omega_T}{\Delta \omega}$.
Outside this width $W$, the LL states can be viewed as boundary modes
with a fixed helicity.
Again each branch of LL that cut the Fermi surface will contribute
a helical surface Fermi surface, and thus the $\mathbb{Z}_2$ topology
does not change, only the Fermi velocities of surface modes
are affected.

\section{Delocalized helical modes on the side boundary}
In the main text, we have solved the helical surface states of the 3D LL
Hamiltonian of Eq. (1) on the $xy$ plane.
Each LL below the chemical potential contributes one helical Fermi surface.
We further check the side surface below.
According to the theorem proved in Ref. [\onlinecite{jiang2011}], as long
as one surface exhibits a helical Fermi surface, all surfaces are
topologically equivalent to odd numbers of channels of gapless
delocalized helical modes.
This theorem also applies to the side surface in our case in which
the translational symmetry is absent along the $z$ direction.
Following the reasoning in Ref. [\onlinecite{jiang2011}], let us
consider two orthogonal surfaces of the $xy$ plane (top) and $xz$ plane
(side) which intersect at the line of $x$ axis.
Without loss of generality, we only consider the gap between the lowest
and the second LLs.
The translational symmetry along the $x$ axis is still maintained in spite of
the open boundaries, and $k_x$ is conserved across the boundary
between $xy$ and $xz$ surfaces.
For a given energy $E$ lying inside this gap, it cuts surface states
on the $xy$ plane with one helical iso-energy ring in momentum space
centering around the origin $(k_x,k_y)=(0,0)$.
Let us focus on the channel of modes with $k_x=0$.
There are a pair of modes $(0,\pm k_y(E))$ on the $xy$ surface, which
cannot be scattered into each other at the intersection edge
between $xy$ and $xz$ surfaces due to TR symmetry.
Each of them has to continue on the $xz$ surface.
According to the theorem in Ref. [\onlinecite{jiang2011}], the
oddness of the number of TR pairs does not change on the $xz$
surface and cannot be localized even though the $z$ axis is
not translational invariant.
With varying the energy $E$ across the gap, we arrive at a branch
of delocalized helical modes on the side surface of the $xz$ plane.
On the other hand, for the case of $k_x\neq 0$, the modes
in the $xy$ surface $(k_x, \pm k_y^\prime)$ with the energy $E$ are not
TR partners, and thus they can be scattered into each other
at the intersection edge.
There is no guaranty that they need to continue on the side
surface of $xz$ plane, or, the modes on the $xz$ plane with
$k_x\neq 0$ are generally gapped and may be localized.
In summary, on the side surface, although we cannot define
Dirac cones due to the lack of translational symmetry,
there do exist an odd number of helical delocalized modes.

\section{Laughlin-type wave functions}
These high dimensional LLs may provide a convenient platform for the
further study of high dimensional interacting fractional topological
states.
The construction of the Laughlin-type wave functions for interacting fermions
is difficult for these SO coupled high dimensional LL systems.
Nevertheless, for the 4D case in the magnetic field, the LLL states with both
$n=0$ and $m=0$ are spin-polarized, and their total DOS is finite
as $\rho^{4D}_{n=m=0}=\frac{1}{4\pi^2 l_{so}^2 l_B^2}$.
Even though they are degenerate with other LLL states with
$(n=0, m\neq 0)$, they are favored by repulsive interactions if they
are partially filled.

The Laughlin wavefunction in the Landau gauge for the 2D LLs has been
constructed in Ref. \cite{rezayi1994}.
We generalize it to the 4D case, and define $w=e^{i\frac{x+iy}{L_x}}$
and $v=e^{i\frac{z+iu}{L_z}}$, then LL states are represented as
$w^{h_x}v^{h_z}$ up to a Gaussian factor $e^{-\frac{u^2}{2l_{so}}-\frac{y^2}{2l_B^2}}$.
The Laughlin-type wavefunction can be constructed as
\bea
\Psi(w_1,v_1; ...; w_{N_xN_y}, v_{N_x N_y}) &=& (\det[ w_i^{h_x} v_i^{h_z} ])^q,
\label{eq:laughlin}
\eea
where $q$ is an odd number;
$\det[ w_i^{h_x} v_i^{h_z} ]$ represents the Slater determinant for the
fully filled single particle states $w^{h_x} v^{h_z}$ and $0\le h_x \le N_x-1$
and $0\le h_z \le N_z-1$.
The study of the topological properties of Eq. \ref{eq:laughlin} will
be deferred to a later publication.

\vspace{5mm}
\section{High dimensional LLs of Dirac electrons}
The LL quantization based on 2D Dirac electrons have attracted a great
deal of research attention since the discovery of graphene.
Here we generalize LLs to the 3D and 4D relativisitc fermions, which are
the square root problems to their Schr\"odinger versions in Eq. 1 and Eq. 6
For the 3D case, we have
\bea
H^{3D}_{LL,Dirac}=\frac{l_{so} \omega_{so}}{\sqrt{2}} \left[
\begin{array}{cc}
 0 & \vec \sigma \cdot \vec p
                         +i \frac{z\hbar}{l_{so}^2} \sigma_z  \\
\vec \sigma \cdot \vec p -i \frac{z\hbar}{l_{so}^2} \sigma_z &0
\end{array}
\right].
\label{eq:3D_Dirac}
\eea
Its square exhibits a diagonal block form with a super-symmetric
structure as
\bea
\frac{(H^{3D}_{Dirac})^2}{\hbar \omega_{so}} =
\left( \begin{array}{cc}
H^{3D,+}_{LL}-\frac{\hbar\omega_{so}}{2}& 0 \\
0&H^{3D,-}_{LL}+\frac{\hbar\omega_{so}}{2}\\
\end{array}
\right),
\eea
where $H^{3D,+}_{LL}$ is just the 3D LL Hamiltonian given in
Eq. 1 in the main text, and $H^{3D,-}$ is given as
\bea
H^{3D,-}_{LL}
&=& \frac{\vec p^2}{2m}+\frac{1}{2}m \omega_{so}^2 z^2
+\omega_{so} z (p_x \sigma_y-p_y \sigma_x).
\eea
The eigenvalues of Eq. (\ref{eq:3D_Dirac}) are $E_{\pm n}=\pm\sqrt {n}
\hbar \omega_{so}$.
The eigen-wave-functions are constructed based on the eigenstates of $H^{3D,\nu=\pm}_{LL}$ as
\bea
\Psi^{3D}_{\pm n,Dirac}(\vec k_{2D},\Sigma)=
\frac{1}{\sqrt 2}
\left(
\begin{array}{c}
\Psi_{n,\vec k_{2D},\Sigma}^+ \\
\pm \Psi_{n-1,\vec k_{2D},-\Sigma}^-
\end{array}
\right)
\eea
where the eigenstate of  $H^{3D,-}_{LL}$ takes the form as
\bea
\Psi^{-}_{n, \vec k_{2D},\Sigma} = e^{i \vec k_{2D} \cdot \vec r_{2D}}
\phi_n [z + z_0( k_{2D},\Sigma)] \otimes \chi_{\Sigma}(\hat k_{2D}).
\eea
The 0th LL states are Jackiw-Rebbi half-fermion modes with only the upper
two components nonzero \cite{jackiw1976,niemi1986}.

The 4D LL Hamiltonian for Dirac Hamiltonian can be constructed as
\bea
H^{4D}_{LL,Dirac}=\frac{l_{so} \omega_{so}}{\sqrt{2}} \left[
\begin{array}{cc}
 0 & \vec \sigma \cdot \vec p_{3D}
                         -i\frac{a_u}{l_{so}}  \\
\vec \sigma \cdot \vec p_{3D} +i \frac{a_u^\dagger}{l_{so}} &0
\end{array}
\right], \ \ \
\label{eq:4D_Dirac}
\eea
where $a_u=\frac{1}{\sqrt 2 l_{so}} (u + i \frac{l^2_{so}}{\hbar} p_u)$
is the phonon annihilation operator in the $u$ direction.
The eigenvalues are still $E_{\pm n}=\pm \sqrt{n}\hbar \omega_{so}$, and
the eigenstates are
\bea
\Psi^{4D}_{\pm n,Dirac}(\vec k_{3D},\Sigma)=\frac{1}{\sqrt 2}
\left ( \begin{array}{c}
\Psi_{n,\vec k_{3D},\Sigma} \\
\pm \Psi_{n-1,\vec k_{3D},\Sigma}
\end{array}
\right).
\eea


\end{document}